\documentclass[12pt,preprint,sort&compress]{elsarticle}
\usepackage{latexsym,amsmath,amsfonts,amssymb}
\usepackage{epsfig,graphics}

\newcommand{\be}{\begin{equation}}
\newcommand{\ee}{\end{equation}}

\newlength{\figsize}
\journal{Phys Lett B}

\begin{document}


\begin{frontmatter}

\title{Mass anomalous dimension in SU(2) with six fundamental fermions}

\author[Jesus]{Francis Bursa}
\author[Edinburgh]{Luigi Del Debbio, Liam Keegan}
\author[Odense]{Claudio Pica} 
\author[Oxford]{Thomas Pickup}
\address[Jesus]{Jesus College, Cambridge. CB5 8BL, United Kingdom}
\address[Edinburgh]{SUPA, School of Astrophysics and Astronomy,
  University of Edinburgh, Edinburgh, EH9 3JZ, United~Kingdom}
\address[Odense]{$CP^3$-Origins, University of Southern Denmark Odense, 5230 M, Denmark}
\address[Oxford]{Rudolf Peierls Centre for Theoretical Physics,
  University of Oxford, Oxford, OX1 3NP, United~Kingdom}

\begin{abstract}

We simulate $SU(2)$ gauge theory with six massless
fundamental Dirac fermions. We measure the running of the coupling and
the mass in the Schr\"{o}dinger Functional scheme. We observe
very slow running of the coupling constant. We measure the mass
anomalous dimension $\gamma$, and find it is between 0.135 and 1.03 in
the range of couplings consistent with the existence of an IR fixed
point.

\end{abstract}

\end{frontmatter}

CP3-Origins-2010-28

DAMTP-2010-51 

\setcounter{page}{1}
\newpage
\pagestyle{plain}

\section{Introduction}
\label{sec:intro}

Phenomenologically viable models of technicolor can be built that are
based on the existence of gauge theories with an infrared fixed point
(IRFP)~\cite{Caswell:1974gg,Banks:1981nn}. The latter are
asymptotically free theories where the dynamical effects of the
fermion determinant induce a non-trivial zero of the beta functions at
low energies, leading to scale-invariance at large distances. In
particular the existence of a large mass anomalous dimension at the
fixed point has been advocated as an important ingredient for
model-building.

The phenomenology of strongly-interacting electroweak symmetry
breaking has been summarized in several interesting reviews (see
e.g. Ref.~\cite{Hill:2002ap} for early results, and
Refs.~\cite{Sannino:2009za,Piai:2010ma} for the more recent
developments.).

Several candidate theories have been singled out by analytical studies
based on approximations of the full nonperturbative dynamics. A
putative phase diagram which summarizes nicely the most appealing
options was discussed in Ref.~\cite{Dietrich:2006cm}, see
Ref.~\cite{Sannino:2008ha} for a review of recent results. These
seminal results have triggered a number of numerical
studies. Numerical simulations of theories defined on a spacetime
lattice are indeed a privileged tool to study the nonperturbative
dynamics of these theories from first principles. Current studies have
focused on SU(2) with two adjoint Dirac
fermions~\cite{Catterall:2007yx,DelDebbio:2008wb,DelDebbio:2008zf,Catterall:2008qk,DelDebbio:2008tv,Hietanen:2008mr,Hietanen:2008vc,DelDebbio:2009fd,Pica:2009hc,Hietanen:2009az,Bursa:2009we},
SU(3) with 8,10,12 fermions in the fundamental
representation~\cite{Fodor:2008hn,Deuzeman:2008sc,Deuzeman:2008pf,Deuzeman:2008da,Deuzeman:2009mh,Deuzeman:2009ze,Fodor:2009rb,Fodor:2009ff,Jin:2008rc,Fodor:2009nh,Fodor:2009wk,Appelquist:2007hu,Fleming:2008gy,Appelquist:2009ty},
SU(3) with two sextet
fermions~\cite{DeGrand:2008dh,DeGrand:2008kx,Fodor:2008hm,Fodor:2009ar,Fodor:2009nh,DeGrand:2009et,DeGrand:2009hu,Svetitsky:2008bw,Shamir:2008pb,DeGrand:2010na},
using a variety of methods. These studies have already revealed an
interesting pattern of results about the phase diagram of
strongly-interacting gauge theories, which can provide useful
comparisons with (and hence guidance for) the analytical results.

In this work we focus on an SU(2) gauge theory with six flavors of
Dirac fermions in the fundamental representation, which is supposed to
be close to the lower boundary of the conformal window for the SU(2)
color group.  For example, the ladder approximation predicts that the
conformal window begins at
$N_f=7\frac{73}{85}$~\cite{Dietrich:2006cm}, the all-orders beta
function conjecture predicts the lower boundary of the conformal
window is at 
$N_f = 22/(2+\gamma_\star)$, which
depends on the fixed point value of 
$0\le\gamma_\star\le2$~\cite{Ryttov:2007cx,Sannino:2009aw},
and metric confinement predicts the conformal window begins at 
$N_f=6.5$~\cite{Frandsen:2010ej}. This model
is particularly relevant as it may be used for a practical realization
of conformal technicolor theories~\cite{Luty:2004ye,Fukano:2010yv}.

Using the Schr\"odinger Functional (SF) formulation of the
theory~\cite{Luscher:1991wu,Luscher:1992an}, we compute the running
coupling and fermion mass as a function of the energy scale in the SF
scheme, thereby deriving information on the beta function and the mass
anomalous dimension. We find clear evidence that the running of the
coupling is rather slow, and indeed compatible with the existence of a
fixed point. As pointed out in previous studies~\cite{Bursa:2009we},
the identification of a fixed point by numerical techniques is
intrinsically difficult: in the vicinity of the fixed point the
coupling changes very slowly as a function of the scale; in order to
detect a slow running, and to be able to identify precisely the
location of the fixed point, great care must be exercised in taming
the systematic errors that arise in numerical simulations. In
particular we need to assess critically the systematic errors that are
involved in our actual procedure, and their propagation in the
analysis of the lattice data.

Contrary to the case of the gauge coupling, the running of the fermion
mass does not slow down at the fixed point, and can be more easily
identified by numerical methods. Results for the step-scaling function
for the scale dependence of the renormalized mass, $\sigma_P$, yield a
bound on the mass anomalous dimension at the fixed point,
$\gamma_*$. Currently the main source of error in the determination of
$\gamma_*$ comes from the uncertainty in the value of the gauge
coupling at the fixed point. First results for the anomalous dimension
were obtained in Ref.~\cite{Bursa:2009we} for the SU(2) gauge theory
with adjoint fermions. Recently new results have been obtained for the
SU(3) gauge theory with sextet fermions in Ref.~\cite{DeGrand:2010na}.

The method used and the observables considered in this work are the
same as the ones we implemented in Ref.~\cite{Bursa:2009we}. They are briefly
summarized for completeness in Section~\ref{sec:SF}, together with the
parameters of the runs that have been used for this analysis. The
running of the coupling is encoded in the step scaling function
$\sigma(u)$; our results for the latter are presented and critically
discussed in Section~\ref{sec:SFcoupling}. Finally the running of the
mass is studied in Section~\ref{sec:SFmass}; the data for the mass
step scaling function $\sigma_P(u)$ compare favourably with the
one-loop perturbative prediction, a feature that we also observed in
our study of the SU(2) gauge theory with adjoint fermions. Even though
we are unable determine whether a fixed point is present, our data are
sufficiently precise to yield an upper bound on the value of the
anomalous dimension throughout the range of couplings that we measure.

\section{SF formulation}
\label{sec:SF}

We measure the running coupling using the Schr\"{o}dinger Functional
method~\cite{Luscher:1991wu,Luscher:1992an}. We follow the same
procedure as in Ref.~\cite{Bursa:2009we} except for the change
from two flavours of adjoint fermions to six flavours of fundamental
fermions. Here we briefly describe the method; for a full description
see Ref.~\cite{Bursa:2009we}.

The Schr\"{o}dinger Functional coupling is defined on a hypercubic
lattice of size $L$. The boundary conditions are chosen to impose  a
constant background chromoelectric field, and depend on a parameter
$\eta$. The coupling constant is then defined as
\begin{equation}
  \label{eq:SFcoupling}
  \overline{g}^2=k \left< \frac{\partial S}{\partial \eta} \right>^{-1}
\end{equation}
with $k=-24L^2/a^2 \mathrm{sin}(a^2/L^2 (\pi-2\eta))$ chosen such that
$\overline{g}^2=g_0^2$ to leading order in perturbation theory. This
gives a non--perturbative definition of the coupling which depends
only on $L$ and the lattice spacing $a$. We then remove the lattice
spacing dependence by taking the continuum limit.

We determine the mass anomalous dimension $\gamma$ from the
  scale dependence of the pseudoscalar density renormalisation
constant $Z_P$. This is defined as a ratio of correlation functions:
\begin{equation}
  \label{eq:ZPdef}
  Z_P(L)=\sqrt{3 f_1}/f_P(L/2)\, ,
\end{equation}
as in Ref.~\cite{Bursa:2009we}.

We use the Wilson plaquette gauge action, together with fundamental
Wilson fermions, and an RHMC algorithm with 4 pseudofermions.

We run at $\kappa_c$, defined as the value of $\kappa$ for which the
PCAC mass $am$ vanishes. We measure $am$ for 5 values of
$\kappa$ for each $\beta$ on $L=6,8,10,12$ lattices and interpolate to find $\kappa_c$ for each.
We then extrapolate in $a/L$ to determine $\kappa_c$ for the $L=14,16$ lattices.

In practice we achieve $|am|\lesssim 0.005$. At some values of $\beta$
and $L$ we have additional results at moderately small masses of
$|am|\sim 0.01$, and we observe no mass-dependence within our
statistical errors, confirming that any residual finite-mass errors
are extremely small.

\subsection{Lattice parameters}
We have performed two sets of simulations in order to determine
$\overline{g}^2$ and $Z_P$. We use more values of $L$ (six instead of
four) compared to our previous simulations to improve the quality of
the continuum limit extrapolations, and increase the step scaling factor
from $s=4/3$ to $s=3/2$ to improve the measurement of the slow running of the coupling.

To ensure our results are not affected by the presence of a bulk
transition, we have measured the average plaquette for a range of
values of $\beta$ and $\kappa$
on $6^4$ lattices with SF boundary conditions.
There is a clear jump in the plaquette at low $\beta$,
suggesting the presence of a bulk transition. However, this disappears
around $\beta=1.6$. Since the lowest $\beta$ we use for our
measurements of $\overline{g}^2$ and $Z_P$ is $\beta=2.0$, our results
should not be affected by this transition.

The parameters of the runs are summarised in Table~\ref{tab:par}. The values
of $\kappa_c$ are obtained from the PCAC relation as described above.

\begin{table}
  \centering

{\footnotesize
  \begin{tabular}{|c|c|c|c|c|c|c|}
    \hline
    $\beta$ & $L$=6 & $L$=8 & $L=10$ & $L$=12 & $L=14$ & $L$=16\\
    \hline
    2.0 & 0.151788 & 0.150970 & 0.150576 & 0.150491 & 0.150334 & 0.150252 \\
    2.2 & 0.147447 & 0.146939 & 0.146755 & 0.146782 & 0.146615 & 0.146565 \\
    2.5 & 0.143209 & 0.142825 & 0.142767 & 0.142811 & 0.142730 & 0.142716 \\
    3.0 & 0.138869 & 0.138684 & 0.138651 & 0.138562 & 0.138523 & 0.138493 \\
    3.5 & 0.136130 & 0.136143 & 0.136104 & 0.136103 & 0.136096 & 0.136091 \\
    4.0 & 0.134394 & 0.134350 & 0.134353 & 0.134339 & 0.134332 & 0.134327 \\
    5.0 & -	   & 0.132142 & 0.132142 & 0.132142 & 0.132142 & 0.132142 \\
    6.0 & 0.130753 & 0.130737 & 0.130748 & 0.130740 & 0.130739 & 0.130738 \\
    8.0 & 0.129131 & 0.129145 & 0.129167 & 0.129172 & 0.129177 & 0.129182 \\
    \hline
  \end{tabular}
}

\caption{Values of $\beta$, $L$, $\kappa$ used for the determination of
  $\overline{g}^2$ and $Z_P$. The entries in the table are the values 
  of $\kappa_c$ used for each
  combination of $\beta$ and $L$.
}

\label{tab:par}
\end{table}

\section{Results for the coupling}
\label{sec:SFcoupling}

We have measured the coupling in the Schr\"{o}dinger Functional
scheme, $\overline{g}^2(\beta,L)$, for a range of
$\beta,L$. Our results are shown in Table~\ref{tab:SFdata}.
We see immediately that the coupling is very similar for different
$L/a$ at constant $\beta$, so it runs slowly.

\begin{table}
  \centering

{\footnotesize
  \begin{tabular}{|c|c|c|c|c|c|c|}
    \hline
    $\beta$ & $L$=6 & $L$=8 & $L=10$ & $L$=12 & $L=14$ & $L$=16\\
    \hline
2.0 & 4.941(61)	&	5.521(143)	&	6.053(418)	&	6.109(289)	&	5.913(362)	&	5.726(485) \\
2.2 & 3.755(32)	&	4.025(70)	&	4.390(158)	&	4.506(345)	&	4.279(233)	&	4.379(252)\\
2.5 & 2.973(21)	&	3.038(37)	&	3.103(72)	&	3.170(67)	&	3.187(174)	&	3.316(151)\\
3.0 & 2.123(10)	&	2.173(20)	&	2.150(37)	&	2.291(90)	&	2.336(55)	&	2.338(75)\\
3.5 & 1.660(8)	&	1.707(37)	&	1.730(20)	&	1.751(29)	&	1.825(50)	&	1.715(46)\\
4.0 & 1.376(4)	&	1.390(8)	&	1.425(16)	&	1.399(30)	&	1.420(19)	&	1.445(31)\\
5.0 & - 	& 	1.033(3)	&	1.054(7)	&	1.050(9)	&	1.063(15)	&	1.041(16)\\
6.0 & 0.814(1)	&	0.822(3)	&	0.823(6)	&	0.842(6)	&	0.829(12)	&	0.827(11)\\
8.0 & 0.576(1)	&	0.581(1)	&	0.575(3)	&	0.586(3)	&	0.585(6)	&	0.593(6)\\
    \hline
  \end{tabular}
}

\caption{The entries in the table are the 
measured values of $\overline{g}^2$ for each
  combination of $\beta$ and $L$. 
}

\label{tab:SFdata}
\end{table}

To analyse the running of the coupling we first define the lattice
step-scaling function,
\begin{equation}
  \Sigma(u,s,a/L) = \overline{g}^2(g_0,sL/a) |_{\overline{g}^2(g_0,L/a)=u}
\end{equation}
and its continuum limit:
\begin{equation}
  \sigma(u,s) = \lim_{a/L\to 0} \Sigma(u,s,a/L)\, ,
\end{equation}
where in both cases we will use only $s=3/2$. We calculate
$\Sigma(u,s,a/L)$ from our data as follows:

We first discard the $L=6$ data since we have found it has large
lattice artifacts. We then interpolate the remaining data
quadratically in $a/L$ at each $\beta$ to find
$\overline{g}^2(\beta,L)$ at $L=9\frac{1}{3},10\frac{2}{3},15$. Then
for each $L$ we interpolate in $\beta$ using the functional form~\cite{Appelquist:2009ty,Bursa:2009we}
\begin{equation}
\label{eq:fitg}
\frac{1}{\overline{g}^2(\beta,L/a)} = 
\frac{\beta}{2N}\left[\sum_{i=0}^{n} c_{i} \left(\frac{2N}{\beta}\right)^{i} \right].
\end{equation}
We choose the smallest $n$ which results in a $\chi^2$ such that the
fit is not ruled out at a $95\%$ CL, and also use $n+1$ as the next
best fit; this gives a 2-5 parameter fit in each case. The number of
parameters we use for each $L/a$ and the $\chi^2/\mathrm{dof}$ for
each fit are shown in Tables~\ref{tab:gfit1} and~\ref{tab:gfit2}.

We now calculate $\Sigma(u,s,a/L)$ using the fits from
Eq.~\ref{eq:fitg} for $L=8,9\frac{1}{3},10, 10\frac{2}{3}$ and
$s=3/2$. Finally we extrapolate to $a/L=0$ to obtain
$\sigma(u,s)$.
Note that this extrapolation only makes sense when we
are on the weak-coupling side of any IRFP.

\begin{figure}
  \centering
  \epsfig{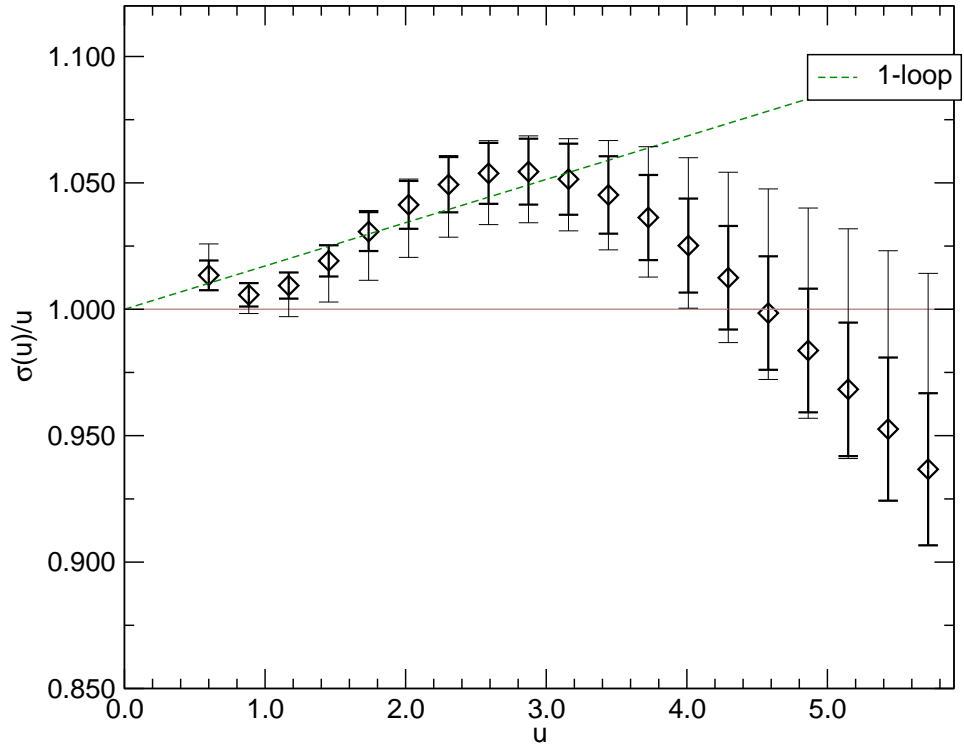}
  \caption{$\sigma(u)$ using a constant continuum extrapolation of the
    two points closest to the continuum. Statistical error using
    the optimal fit parameters in black, systematic error
    from using different numbers of parameters in the
    fits in grey.}
  \label{fig:SFsigcon}
\end{figure}

We carry out a constant continuum extrapolation, using the data at the
two values of $a/L$ closest to the continuum limit. We estimate the
errors using a multistage bootstrapping procedure. 
The results for $\sigma(u)$ using the constant continuum
extrapolation are plotted in Figure~\ref{fig:SFsigcon}, where the
statistical errors only are in black and the error from changing the
number of fitting parameters are in grey. Our results are
consistent with a fixed point in the region
$\overline{g}^2 > 4.02$. They are also
compatible with the possibility that there is no fixed point at all in
the range of couplings we have measured. However, it is clear that $\sigma(u)$ is
considerably below the 1-loop prediction at strong coupling.

We have also carried out a linear continuum extrapolation, using the
data at the four values of $a/L$ closest to the continuum limit, but it is
not as well constrained by our data. However, it is also below the
1-loop prediction at strong coupling, and has $\sigma(u)/u < 1$ in the region
$\overline{g}^2 > 4.08$. This is consistent with the presence of a
fixed point in the region $\overline{g}^2 < 4.08$.

There is a large difference between the constant and continuum
extrapolations, showing that the systematic error from the choice of continuum
extrapolation is large. This means we cannot determine conclusively whether or
not there is a fixed point in the range of couplings covered by our data.

\begin{table}
\centering
{\footnotesize
\begin{tabular}{|c||c|c|c|c|c|c|c|c|}
\hline
$\overline{g}^2$ & \multicolumn{8}{|c|}{$L/a$} \\
\hline
& 8 & $9\frac{1}{3}$ & 10 & $10\frac{2}{3}$ & 12 & 14 & 15 & 16 \\
\hline
params & 3 & 4 & 4 & 4 & 3 & 2 & 2 & 2 \\
$\chi^2/\mathrm{dof}$ & 1.92 & 0.54 & 1.24 & 0.48 & 1.66 & 1.54 & 1.88 & 1.36 \\
\hline
\end{tabular}
}
\caption{Interpolation best fit parameters for $\overline{g}^2$.
}
\label{tab:gfit1}
\end{table}

\begin{table}
\centering
{\footnotesize
\begin{tabular}{|c||c|c|c|c|c|c|c|c|}
\hline
$\overline{g}^2$ & \multicolumn{8}{|c|}{$L/a$} \\
\hline
& 8 & $9\frac{1}{3}$ & 10 & $10\frac{2}{3}$ & 12 & 14 & 15 & 16 \\
\hline
params & 4 & 5 & 5 & 5 & 4 & 3 & 3 & 3 \\
$\chi^2/\mathrm{dof}$ & 1.25 & 0.58 & 1.42 & 0.54 & 1.19 & 1.61 & 1.06 & 1.05 \\
\hline
\end{tabular}
}
\caption{Interpolation next-best fit parameters for $\overline{g}^2$.
}
\label{tab:gfit2}
\end{table}

In the vicinity of a fixed point at a coupling $g^*$, the beta
function is linear in the coupling,
\begin{equation}
\beta(g) = \beta^* (g-g^*) + ... 
\end{equation}
where $\beta^*$ is a scheme-independent coefficient, which, as
described in Ref.~\cite{Gardi:1998ch}, yields further information on
the physics of these theories.  In terms of the
step-scaling function $\sigma(u,s)$, this gives:
\begin{equation}
\sqrt\sigma(u,s) = g^* + (\sqrt u - g^*) s^{-\beta^*}.
\label{sigma_beta*}
\end{equation}
We have estimated $\beta^*$ by fitting $\sigma(u,s)$ in the vicinity of 
the fixed point, and find $\beta^*=0.62(12)^{+13}_{-28}$, where the 
first error is statistical and the second is systematic, for those fits 
where we see a fixed point in the range of couplings covered by our 
data. This does not include the systematic error due to the choice of a 
constant rather than a linear continuum extrapolation. Better data 
would be needed to make the systematic errors on $\beta^*$ more robust. 


\section{Running mass}
\label{sec:SFmass}

\begin{figure}
  \centering
  \epsfig{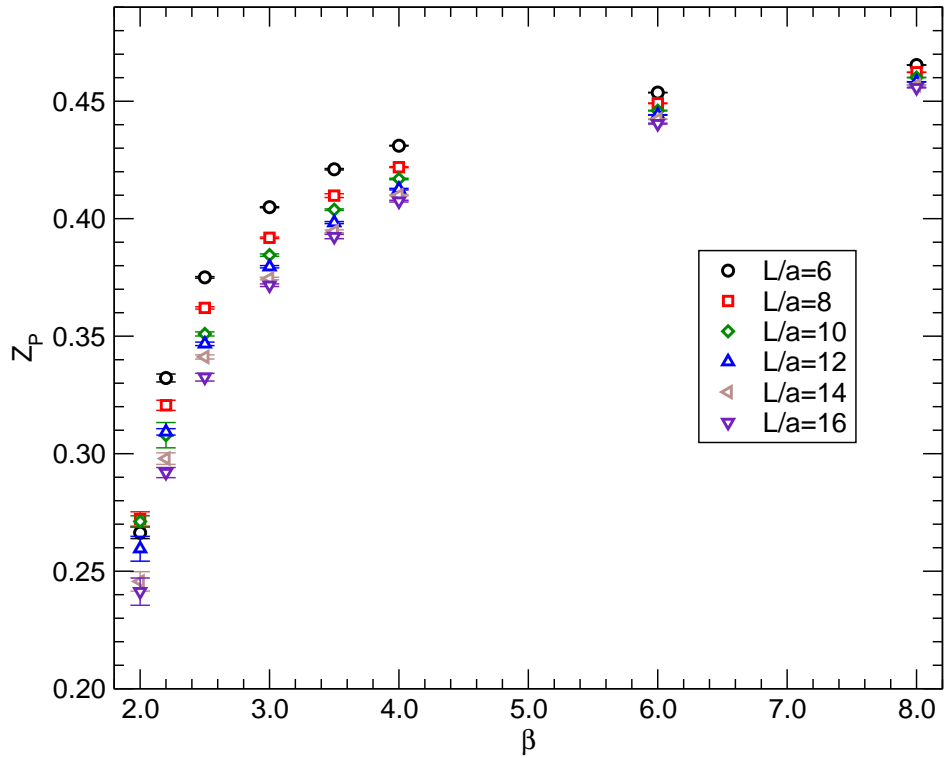}  
  \caption{Data for the renormalisation constant $Z_P$ as computed
    from lattice simulations of the Schr\"odinger
    functional. Numerical simulations are performed at several values
    of the bare coupling $\beta$, and for several lattice resolutions
    $L/a$.}
  \label{fig:ZPdata}
\end{figure}

We have measured the pseudoscalar density renormalisation constant
$Z_P$ for a range of $\beta,L$. Our results are shown in Table~\ref{tab:ZPdata} and Figure~\ref{fig:ZPdata}.
We see that $Z_P$ decreases with increasing
$L/a$ at constant $\beta$, indicating a positive anomalous mass
dimension, but the running appears to be slow.

\begin{table}
  \centering

{\footnotesize
  \begin{tabular}{|c|c|c|c|c|c|c|}
    \hline
    $\beta$ & $L$=6 & $L$=8 & $L=10$ & $L$=12 & $L=14$ & $L$=16\\
    \hline
2.00 & 0.26636(249)	&	0.27219(306)	&	0.27117(241)	&	0.25956(527)	&	0.24564(414)	&	0.24130(578)\\
2.20 & 0.33220(167)	&	0.32060(216)	&	0.30788(537)	&	0.30929(137)	&	0.29792(246)	&	0.29198(215)\\
2.50 & 0.37504(32)	&	0.36203(49)	&	0.35095(87)	&	0.34672(73)	&	0.34118(88)	&	0.33255(162)\\
3.00 & 0.40488(21)	&	0.39186(31)	&	0.38451(52)	&	0.37955(50)	&	0.37453(53)	&	0.37170(56)\\
3.50 & 0.42102(30)	&	0.40981(82)	&	0.40383(32)	&	0.39832(43)	&	0.39461(62)	&	0.39241(93)\\
4.00 & 0.43105(14)	&	0.42192(21)	&	0.41691(31)	&	0.41256(34)	&	0.40997(29)	&	0.40746(36)\\
6.00 & 0.45368(8)	&	0.44908(12)	&	0.44597(16)	&	0.44417(10)	&	0.44232(15)	&	0.44045(20)\\
8.00 & 0.46540(5)	&	0.46229(7)	&	0.46005(10)	&	0.45822(7)	&	0.45683(10)	&	0.45575(9)\\
    \hline
  \end{tabular}
}

\caption{The entries in the table are the 
measured values of $Z_P$ for each
  combination of $\beta$ and $L$. 
}

\label{tab:ZPdata}
\end{table}

To extract $\gamma$ we first define the lattice step-scaling function
for the mass,
\begin{equation}
  \label{eq:sigmaPdef}
  \Sigma_P(u,s,a/L)=\left
    . {\frac{Z_P(g_0,sL/a)}{Z_P(g_0,L/a)}}
  \right|_{\overline{g}^2(g_0,L/a)=u}\, 
\end{equation}
and its continuum limit
\begin{equation}
  \label{eq:sigma_p}
  \sigma_P(u,s) = \lim_{a\to 0}\Sigma_P(u,s,a/L)\, .
\end{equation}

Again, we use only $s=3/2$. To calculate $\Sigma_P(u,s,a/L)$, we
proceed similarly as for $\Sigma(u,s,a/L)$. We first discard the $L=6$
data, and then interpolate quadratically in $a/L$ to find
$Z_P(\beta,L)$ at $L=9\frac{1}{3},10\frac{2}{3},15$.
Then
for each $L$ we interpolate in $\beta$ using the functional form~\cite{Bursa:2009we}
\begin{equation}
\label{eq:fitZ}
Z_{P}(\beta, L/a) = \sum_{i=0}^{n} c_{i}\left(\frac{1}{\beta}\right)^{i}
\end{equation}
We choose the smallest $n$ which results in an acceptable $\chi^2$, as
for the $\overline{g}^2$ fits; this gives a 5-6 parameter fit in each
case.  We also use $n+1$ as a next-best fit to estimate the systematic
errors from the choice of $n$. The number of parameters we use for
each $L/a$ and the $\chi^2/\mathrm{dof}$ for each fit are shown in
Tables~\ref{tab:zpfit1} and~\ref{tab:zpfit2}.

We can now calculate $\Sigma_P(u,s,a/L)$ using
Eq.~\ref{eq:sigmaPdef} and the fits from Eq.~\ref{eq:fitZ},
and finally extrapolate to the continuum limit to obtain
$\sigma_P(u,s)$.
The errors are smaller than for the running coupling, so we are able
to combine both constant and linear continuum extrapolations to control
the systematic error from the choice of extrapolation.
We estimate the errors using a multistage bootstrapping procedure. 

\begin{figure}
  \centering
  \epsfig{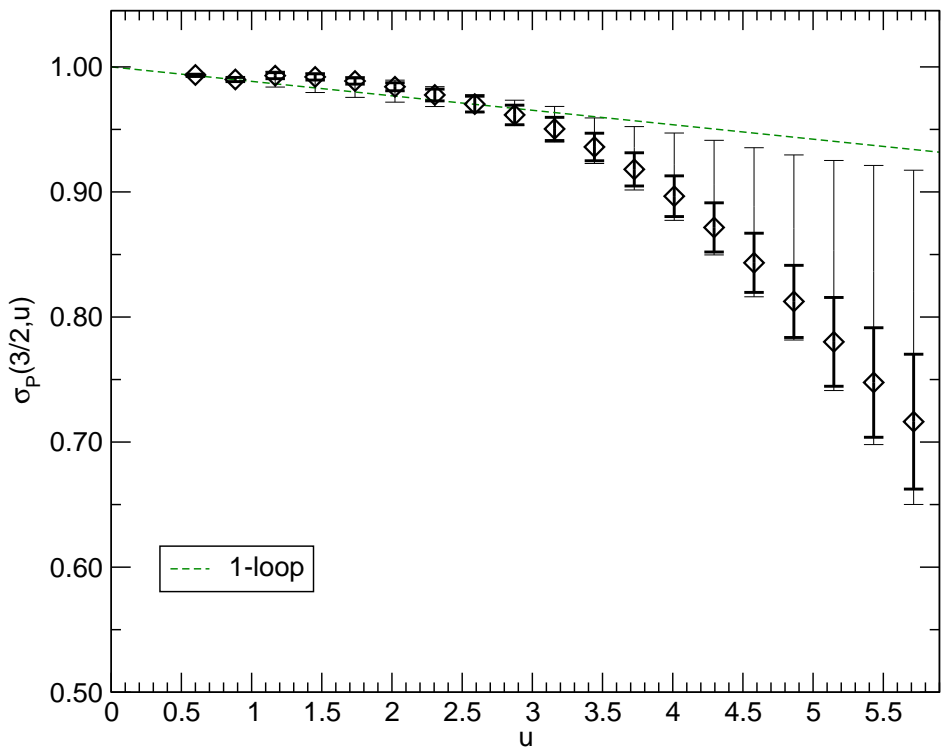}
  \caption{$\sigma_P(u)$ using both a constant continuum extrapolation
    of the two points closest to the continuum, and a linear continuum
    extrapolation. Statistical error using the optimal fit
    parameters with a linear continuum extrapolation in black,
    systematic error including the choice of continuum
    extrapolation in grey.}
  \label{fig:ZPsigboth}
\end{figure}

We plot $\sigma_P$ in Figure~\ref{fig:ZPsigboth}, where the statistical
error is in black, and the systematic error arising both from the
choice of the number of fit parameters and the continuum extrapolation
is in grey. We also plot the 1-loop perturbative prediction for $\sigma_P$.
Our results are close to the 1-loop prediction, with the running
becoming a little faster at strong couplings.

\begin{figure}
  \centering
  \epsfig{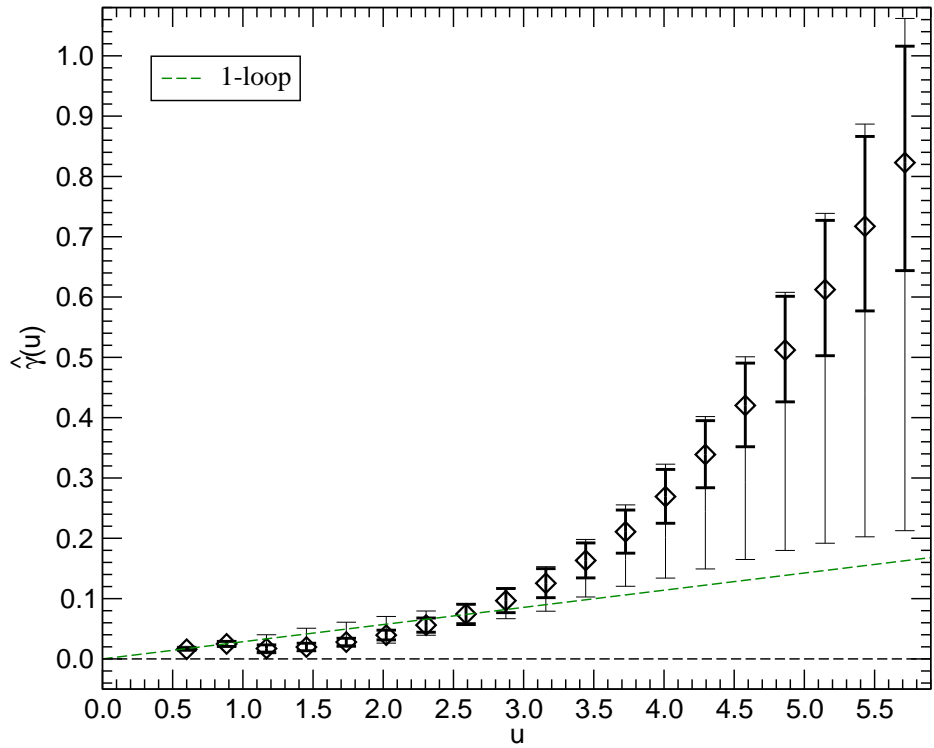}
  \caption{$\hat\gamma(u)$ using both a constant continuum extrapolation of the two points closest to the continuum, and a linear continuum extrapolation. Statistical error using the optimal fit parameters with a linear continuum extrapolation in black,
    systematic error including the choice of continuum
    extrapolation in grey.}
  \label{fig:gamma}
\end{figure}

In the vicinity of an IRFP, we can define an estimator $\hat\gamma(u)$
given by
\begin{equation}
\label{eq:tau}
\hat\gamma(u) =
-\frac{\log\left|\sigma_P(u,s)\right|}{\log\left|s\right|}\, ,
\end{equation}
which is equal to the anomalous dimension $\gamma$ at the fixed
point~\cite{Bursa:2009we}, and deviates away from the fixed point as
the anomalous dimension begins to run. We plot this estimator in
Figure~\ref{fig:gamma}. Again the black error bars show the
statistical errors, and the grey the systematic errors. 
We see that $\hat\gamma(u)$ is small in most of the range of couplings
that we measure. However, it becomes larger at our strongest
couplings. Our data is consistent with it reaching
values $\gamma \approx 1$ that are interesting for models of
technicolor,
although our error bars are large and it is also possible that it is as
small as $0.135$, our lower bound at $\overline{g}^2=4.02$, the
smallest coupling at which a fixed point is consistent with our
results using a constant continuum extrapolation.
The highest value compatible with our data is $\hat\gamma=1.03$ at
$\overline{g}^2=5.52$, the highest coupling at which we have results
for all $L$.

\begin{table}
\centering
{\footnotesize
\begin{tabular}{|c||c|c|c|c|c|c|c|c|}
\hline
$\overline{g}^2$ & \multicolumn{8}{|c|}{$L/a$} \\
\hline
& 8 & $9\frac{1}{3}$ & 10 & $10\frac{2}{3}$ & 12 & 14 & 15 & 16 \\
\hline
params & 6 & 5 & 5 & 5 & 5 & 5 & 5 & 5 \\
$\chi^2/\mathrm{dof}$ & 1.79 & 0.86 & 1.09 & 0.62 & 0.99 & 1.60 & 1.63 & 1.22 \\
\hline
\end{tabular}
}
\caption{Interpolation best fit parameters for $Z_P$.
}
\label{tab:zpfit1}
\end{table}

\begin{table}
\centering
{\footnotesize
\begin{tabular}{|c||c|c|c|c|c|c|c|c|}
\hline
$\overline{g}^2$ & \multicolumn{8}{|c|}{$L/a$} \\
\hline
& 8 & $9\frac{1}{3}$ & 10 & $10\frac{2}{3}$ & 12 & 14 & 15 & 16 \\
\hline
params & 7 & 6 & 6 & 6 & 6 & 6 & 6 & 6 \\
$\chi^2/\mathrm{dof}$ & 2.09 & 0.46 & 1.03 & 0.43 & 1.18 & 0.93 & 1.32 & 1.47 \\
\hline
\end{tabular}
}
\caption{Interpolation next-best fit parameters for $Z_P$.
}
\label{tab:zpfit2}
\end{table}

\section{Conclusions}
\label{sec:concl}

In this work we have calculated the running of the Schr\"{o}dinger
Functional coupling $\overline{g}^2$ and the mass in the continuum
limit of $SU(2)$ lattice gauge theory with six flavours of fundamental
fermions, over a wide range of couplings up to $\overline{g}^2 \approx 5.5$.

Our results for the running of the coupling have relatively large
errors. This is due to the difficulty of measuring the small
difference in the coupling between two nearby scales, a problem that
becomes particularly acute as we approach a possible fixed point where
the difference falls. We observe that the running of the coupling is
slower than the (already slow) one-loop perturbative prediction. Our
results using a constant continuum extrapolation are consistent with the presence of a fixed point above
$\overline{g}^2=4.02$, but it is also possible that there is no fixed
point in the range of couplings we have measured.
There is an additional uncertainty arising from the choice of
continuum extrapolation.

Our results for the running of the mass are clearer. We find the
anomalous dimension is small throughout most of the range of couplings
we measure, but it becomes larger for our strongest couplings, with a
possibility that it reaches values around 1. If true, this would be
very interesting for technicolor models.

The value of $\gamma$ at the fixed point can be predicted using the
all-orders beta function conjecture~\cite{Ryttov:2007cx}. This gives
an exact prediction in terms of group-theoretical factors only. 
If the present theory is inside the conformal window, the prediction
is $\gamma=5/3$, which lies outside the range measured in this study.
Unfortunately, given the uncertainty on the existence of the fixed
point, any conclusion on the validity of the all-order beta function
conjecture is speculative at present.

The accuracy of our results would be improved in particular by using
larger lattices, which would give a larger range of $a/L$ for the
continuum extrapolations. This would help to clarify the existence and
location of the fixed point, and to reduce the errors on the anomalous
dimension. Calculations to improve the statistics and use larger
lattice sizes are ongoing.

\section*{Acknowledgements}
This work was performed using the Darwin Supercomputer of the
University of Cambridge High Performance Computing Service
(http://www.hpc.cam.ac.uk/), provided by Dell Inc. using Strategic
Research Infrastructure Funding from the Higher Education Funding
Council for England; computing resources funded by the University
of Oxford and EPSRC; and the Horseshoe5 cluster at the supercomputing facility at the
University of Southern Denmark (SDU) funded by a grant of the Danish
Centre for Scientific Computing for the project `Origin of Mass'
2008/2009. LDD wishes to thank the Aspen Center for Physics where part
of this work has been carried out. LDD is supported by an STFC
Advanced Fellowship.

\bibliographystyle{model1-num-names}

\vfill\eject

\end{document}